# Comparison of methods of automatic blood pressure measurement in the same device

Sikorskyi M.V., Soroka A.O., Mosiychuk V.S., Sharpan O.B.
The National Technical University of Ukraine "Kyiv Polytechnic Institute"
Radioengineering Faculty
Kyiv, Ukraine
e-mail: mvsikorsky@gmail.com

*Abstract* — **Comparison of tacho-oscillographic and oscillometric methods for human blood pressure measurement with cuff occlusion in the same device at the same time is described. For this purpose the measurement system and software for signal processing that realize these methods in the same device was developed. Experiment shows that oscillometric method with photoplethysmographic (PPG) sensors does not require empiric criteria to search systolic and diastolic blood pressure parameters due to possibility to apply correlation analysis. Perspectives of applying oscillometric method with PPG sensors in automatic blood pressure monitors is showed.**

## I. Introduction

The definition and control of blood pressure (BP) today are considered to be the main preventive measures that preclude the risk of appear and development of cardiovascular diseases and help to avoid mortal consequences of this diseases, which are the main reason of premature death. (More, than in 60 % cases) [1]. Blood pressure is one of the factors of functional state of the organism in general and factors of cardiovascular system functioning in particular. Despite this in automotive BP measuring devices empirical criteria of parameters determination are used, with a help of tacho-oscillographic method with hand cuff [2]. It decreases the authenticities of measurement. We could increase the authenticity of blood pressure definition with the help of multiple measurements, but such method could be precipitated for the sick. Frequent and long-lasting occlusion of the limbs could become the main factor of disturbance and the reason of stress, and also could lead to the puffiness and blood congestion. Non-occlusive (cuff free) methods of the measurement [3] are less precipitated but we could achieve it with the help of decreasing the accuracy and authenticity of measurements.

Also we should mention that almost all existing methods of BP measurements necessitate immovable state of the limb, which participate in measurement. But for actuated artifacts [4] authentic measurements with the help of tacho-oscillographic method in automotive mode become unreal. Further direction of this problem solution is a registry of additional biological signals and their total next digital processing [5].

## II. Problem statement and research purpose

The purpose of the article is comparison of blood pressure measurement methods in one system of automatic measurement at the same time and with the same conditions considering the criteria of authenticity and accuracy measurement and potential resistance to actuated artifacts and insensibility to the personal physiological features of the organism.

## III. Assessment of methods for automatic blood pressure measurement

The methods of blood pressure measurement could be divided in 2 big separate groups:

- Invasive method
- Non-invasive method

Invasive method of BP measurement is used only in hospital conditions during surgical interference, when injection of sensitive element with pressure detector in patient's artery is necessary for permanent control of pressure level. The advantage of this method is that pressure is measuring constantly reflecting in the shape of pressure/time graph. However, patients with invasive monitoring of BP require constant attendance but for hazard of hemorrhage appear in case of sensitive elements disconnecting, formation of hematoma or thrombosis in a place of centesis, occurring of contagious complication. As far as direct method measured pressure inside the vessel, it is absolutely accurate (not considering the allowance of the measurement of the vary equipment) and it is an etalon for assessment of the measurement accuracy of the other methods.

Non-invasive methods are divided in occlusive in which occlusion (cross-clamping) in limb for measurement is needed, in which pressure is measured and non-occlusive. Non-occlusive methods of blood measurement are impossible to being used in automatic mode because of complexity of realization and calibration and that's why they are not considered in the article [3,6].

Auscultatory method (acoustic, Korotkov's tone method) is based on the strepitus registry – Korotkov's tones, that appear in the artery during the decompression of the cuff. The value of the pressure in the cuff at the moment of the first tone registry is defined as systolic pressure (SBP), and at the moment of their disappearance – diastolic pressure (DBP). Auscultatory methodology nowadays is accepted by World Health Organization (WHO) as referential method of non-invasive BP determination, in spite of a slightly undercharged values for SBP and overestimated values – for DBP in comparison to values, which were got in the process of invasive measurement.

The important advantages of the method are a higher resistance to disturbance in the heart and movements of the arm during the measurement. However, there are some essential disadvantages in the method related to high sensitivity to the noises in the premises, bars that appear but for abrasion between the cuff and cloth, and also related to a necessity of accurate placement of the microphone under the artery. Accuracy of BP registry essentially decreases in case of low tones intensity, presence of "auscultatory gap" or "permanent tone". Allowance of BP measurement by this method consists of allowance of vary method, allowance of manometer and moment determination accuracy of reading values.

Tacho-oscillographic method means registry and pulsation pressure analysis – tacho-oscillographic charts (TOG) in compressive cuff. The advantage of this method is absence of any other sensors despite pressure sensor in compressive cuff. The disadvantage of this method is that oscillations of the cuff appear earlier, than blood begins to run through the cuff in the limb. It is explained by the fact that blood flow with every cardiac impulse will cause stronger and stronger pulsation in the cuff. During the pressure decreasing in it, "bitten" in bar and not reaching the remote part of the cuff. It means that pressure value will be overestimated in comparison with the real value. Special processing algorithms are used for this, in which quieting level is established, it is defined by the way of multiple experience and comparisons with more accurate method – acoustic method.

One of the non-invasive method of BP measurement is an oscillometric method with a reference channel (Oscillometric method) [7,8], in which pulse wave – photoplethysmogram are registered and analyzed. Photoplethysmographic method use sensors that are capable to register of absorption by the hemoglobulin of infrared waves, it depends on the pulsing filling tissue with blood. By using the system with reference channel (one detector is placed on the squeezed limb, other- on the free limb), it is possible to measure the values of blood pressure by comparing of photoplethysmographic signals from both sensors and fixing pressure values in the cuff in the moment of fractional coincidence of the pulsing waves (high pressure value) and full coincidence (low BP value). But this method also has disadvantages, for example, necessity of detectors connection to the both limbs. Fractionally, this disadvantage could be improved by using compressing cuff, that is put on the one finger of the arm [7]. In such a case photoplethysmogram of the main and reference channels could be measured on the one limb.

## IV. OSCILLOMETRIC AND TACHO-OSCILLOGRAPHIC METHODS IMPLEMENTATION IN AUTOMATED BLOOD PRESSURE MEASUREMENT DEVICE

System of BP measurement with the registry with additional signals for the possibility of realization of different measurement methodologies in the same one device was developed during this investigation. The structure of this device is shown in Fig.1

The system includes the device for photoplethysmogram registry from both limbs, compressive cuffs with pressure sensor and also software for the processing of received data on PC. The main element of the controlling device is microcontroller with integrated analog digital converter (ADC) and standard USB interface. The device serves three analog measurement channels for BP measurement: two channels with pulse sensors and one channel of pressure sensor in the cuff.

According to the structure pulse channels are same. Microcontroller forms modulated light flow with frequency 1.5 kHz with its further synchronic detecting after running through the biological tissues. Analog highway of signals processing includes transimpedance converter, input amplifier and band filter on operating amplifiers. Analog signals processing includes:

- converting of modulated thermal LED current to the pressure signal and premature amplification of this signal;

- regulation of input signal amplitude to the level when dynamic range of the ADC input are used maximally;

- essential decreasing of common-mode interference that could be put in connective cables;

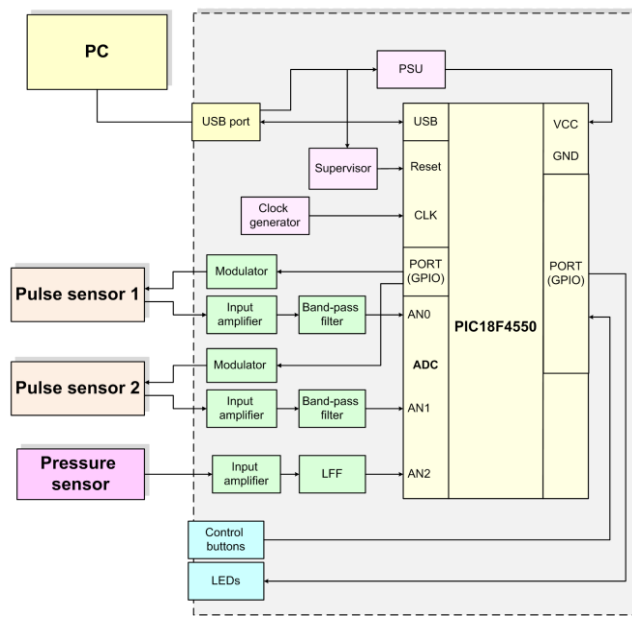

Fig. 1. Structure of blood pressure measurement device

- protection of the system's inputs from potential overload by the pressure input.

From amplifier output the signal is set to the band filter, which has two functions: at first, emphasize only modulated signal (it means that it delete from the signal spectrum high frequency, impulsive and low frequency interferences); secondary, filter is necessary antialliasing filter for further AD conversion. Considering the fact that the speedwork of ADC is much higher than high frequency of spectrum of the pulse signal (30 Hz) and computational resources of processor are enough for premature signal processing, a possibility of signal resampling is used. It means that the sampling frequency could be selected much more bigger than Nykuist frequency for pulse signal spectrum. In that way more affective precision of ADC was achieved, and also it allows to use more simple antialliasing analog filters (for instance second range filter). The channel of pressure sensor in the cuff includes preamplifier and second range LFF.

## V. Signal processing

For BP measurement by oscillometric method the correlative of pulse signal processing of the main and reference channels is realized in the program. According to the absence of occlusion the signals in the main and reference channels are almost identical by the shape and correlative characteristic takes the value ~100%. In the case of full limb occlusion on which BP measurement takes place (pressure in the cuff is higher than systolic pressure), pulsation in the main channel disappear and the value of correlative characteristics ~0%. In the case of fractional occlusion (the pressure in the cuff is lower than systolic but higher than diastolic), pulsing signal of the main channel that is registered lower than cuff, will be transformed and will have a certain shift in time if relation to the signal of the reference channel. During the compression and pressure decreasing in the cuff, the pulsation amplitude in the main channel is increasing, the shape will become more like pulsing shape in the reference channel, and time shift between the main and reference channel disappears. For authenticity of correlation of signal processing increasing, digital filters for interference elimination are realized, as a rule they are in phase and influence on correlative characteristics.

Correlative processing of the signals of PPG from the both channels could be identified with (1), and the distinctive moment in the time $t_{sys}$ and $t_{dyas}$ on the base of analysis of regulated CCF by criteria 10 and 90% accordingly [7]:

$$b(\tau) = \frac{B(\tau)}{\sigma_1 \cdot \sigma_2} = \frac{\text{cov}[S_1(t), S_2(t-\tau)]}{\sigma_1 \cdot \sigma_2} =$$

$$= \frac{\int_{t_0}^{t_0+T-\tau} S_1(t) \cdot S_2^2(t-\tau) dt}{\sqrt{\int_{t_0}^{t_0+T-\tau} S_1^2(t) dt \cdot \int_{t_0}^{t_0+T-\tau} S_2^2(t-\tau) dt}}, \quad (1)$$

where $S_1(t)$, $S_2(t)$ – PPG signals in the main and reference channels;

$\sigma_1$, $\sigma_2$ – root-mean-square deviation of these signals;

$\tau$ – relative time shift of the signals S1 and S2;

$t_0$ – any start moment;

T – the time of integration – duration of time gap

$B(\tau)$ – non-regulated SSF.

For oscillation detection in the cuff and BP measurement by tacho-oscillographic method, data array which were got from the pressure sensor flows through the digital narrow-bent filter with pass frequency 0,5-2 Hz and depression 80 dB out of passing bend

The main window of the program (Fig 2) contains free graphical fields: the first and second – according to the signals from the first and second optical sensors, third graphical field contains the signal from the pressure sensor of compressive cuff (red line), calculated values of correlative function (blue line) and pointed signal of oscillation which appear in the cuff during its decompression (pink line).

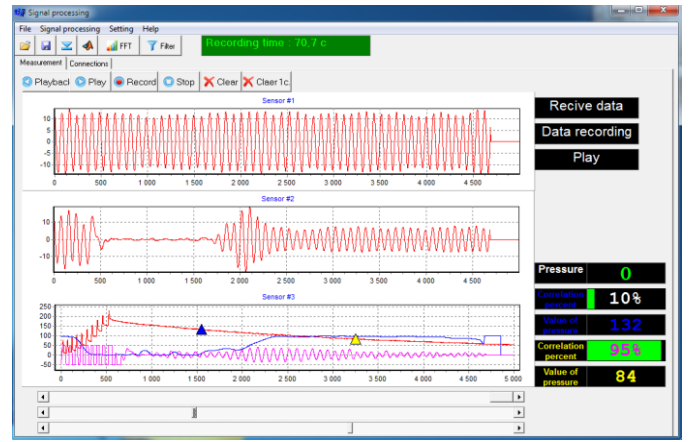

Fig. 2. Software implemented signal processing for blood pressure determination

## VI. Experiment

The purpose of carrying out experiment is the comparison of two measurement methods of blood pressure – tacho-oscillographic and oscillometric with etalon (auscultative) method, using one device that increase the authenticity of comparison. The measurement was performed in comfortable conditions for experimental object. For elimination of the syndrome "white coat" testing measurement were performed (patient calms down with the time). During the experiment this factor was considered, that increases the objectivity of received results. During the preparation for measurement performing all points of BP measurement methodology were considered. They were confirmed in WHO 1999:

- all measurements were carried out in lying position;

- measurement was carried out after two hours after last meal;
- the patient was free of tight cloth;
- arm that fell for occlusion was completely clothless;
- measurements were performed with 5 minutes interval;

*A. Process of experiment.*

The patient is in lying position, after that the cuff is put on one limb and PGG sensors connect to the fingers on different limbs. Software is starting on the PC and stethoscope is placed on the cubital fossa. The air is pumping in the cuff to the value higher than 150 mm.Hg. Air is deflated from the cuff in such a way, so that the speed of pressure dropping in it won't exceed 2 mm.Hg per second. The moment appearance of the first and last tones is registered according to the pressure sensor values of the compressive cuff, which reflect in the real time on the PC screen. After air deflation from the cuff, results of al measurement and all data are saved in PC memory and BP value is determined by the curve of correlative function. The measurements are carried out after again after 5 minutes.

*B. Results and discussion*

During the process of experiment it was picked up ten values of BP from six different people, that displayed on the fig.3.

On the fig.3 axes of abscissa shows the values of systolic pressure and axes of ordinates – diastolic pressure. The points emphasized by round markers are the results of BP by auscultative criss-cross markers are the values of tacho-oscillographic measurement, and the square are the values of oscillometric.

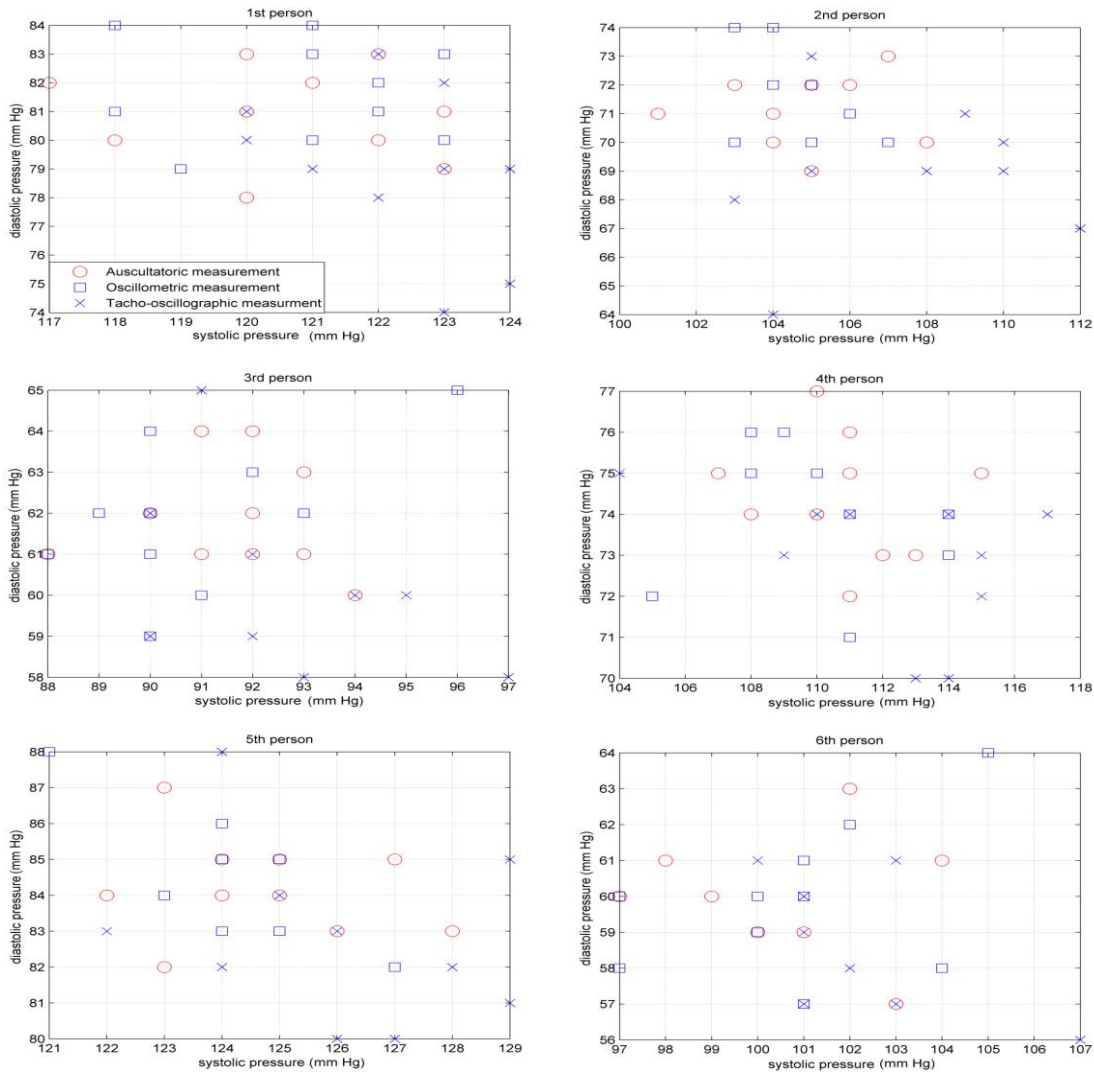

Fig. 3. Results of measuring blood pressure by different methods in the same BP measurement system

Absolute deviation of BP values from auscultative values during the measurement by different methods is shown on fig.4.

The average deviation of BP values that was measured by oscillometric and tacho-oscillographic methods are shown in table 1.

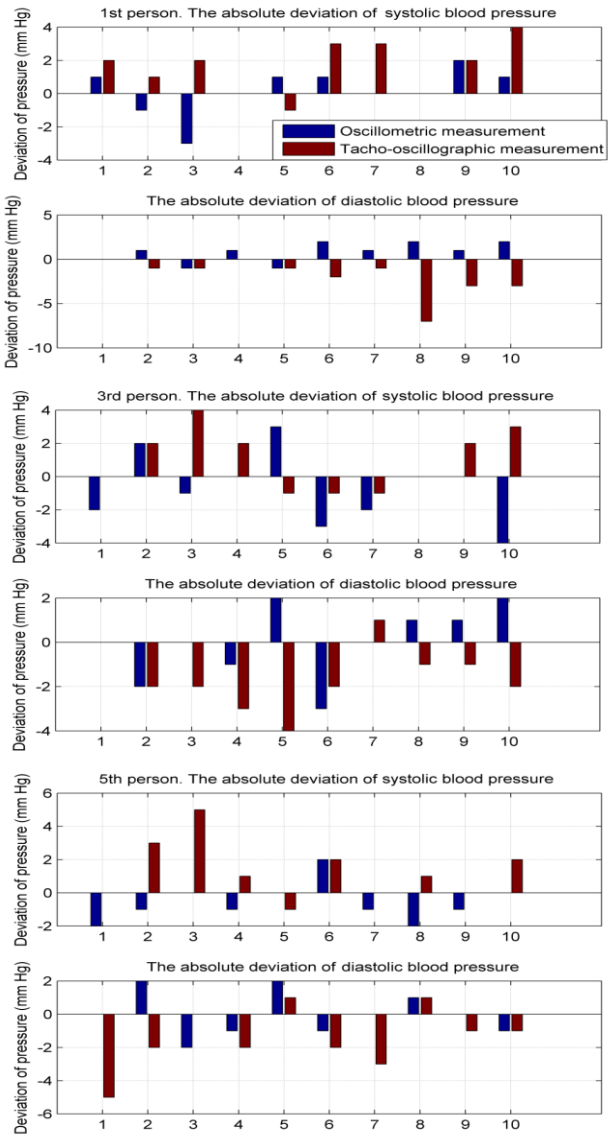
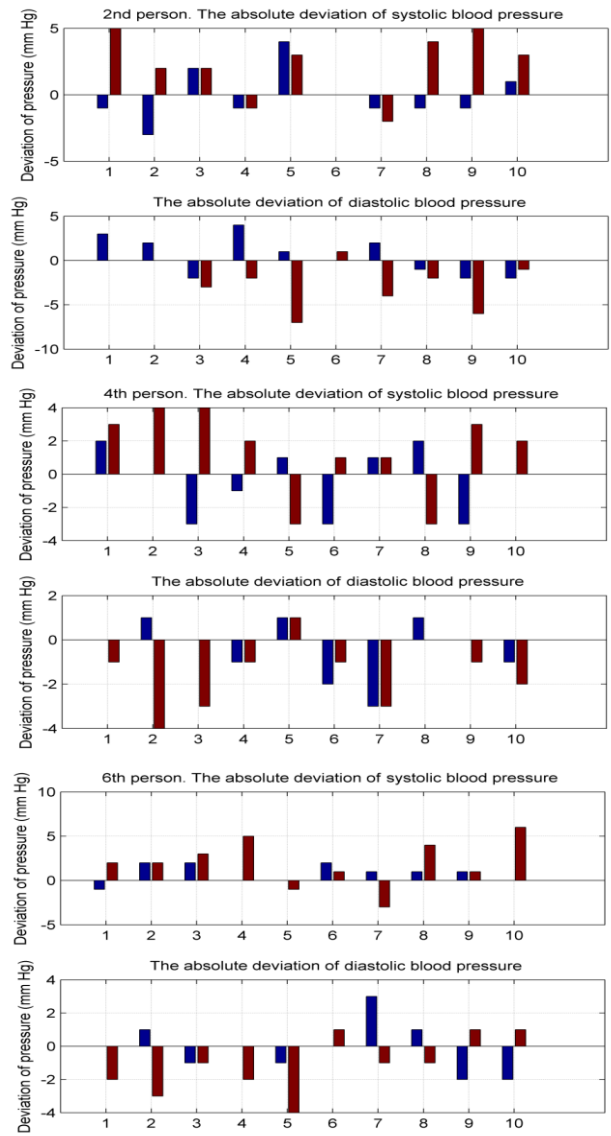

Fig. 4. Absolute deviation of BP values from auscultative

TABLE I. THE AVERAGE DEVIATION OF BP VALUES

| Parameter | Person # | | | | | |
|---|---|---|---|---|---|---|
| | *1* | *2* | *3* | *4* | *5* | *6* |
| P1 | 1.8 | 1.5 | 1.7 | 1.6 | 1 | 1 |
| P2 | 1.2 | 1.9 | 1.2 | 1 | 1 | 1.1 |
| P3 | 1.8 | 2.7 | 1.6 | 2.6 | 1.5 | 2.8 |
| P4 | 1.9 | 2.6 | 1.8 | 1.7 | 1.8 | 1.7 |

P1 - The average deviation of the systolic pressure value received by oscillometric method from value received by auscultative method mm.HG;

P2 - The average deviation of the diastolic pressure value received by oscillometric method from value received by auscultative method mm.HG;

P3 - The average deviation of the systolic pressure value received by tacho-oscillographic method from value received by auscultative method mm.HG;

P4 - The average deviation of the diastolic pressure value received by tacho-oscillographic method from value received by auscultative method mm.HG

This experiment leads that oscillometric and tacho-oscillographic methods of measurement of BP are accurate enough. The average measurement deviation by both mentioned methods doesn't exceed 3 mm.HG. During the measurement tacho-oscillographic method the tendency of overvaluation of systolic pressure is appeared, it explains by existence of oscillation of the cuff before pressure in it decreased to the level when blood flow refreshes in fractionally squeezed artery.

## VII. Conclusion

In hospital conditions BP measurement are carried out synchronically by three methods, that realized in one measurement system, the accuracy of results is approximately same. However, considering the features of signals formation according to which the estimation of parameters of systolic and diastolic pressure is performed, we can claim that with ambulatory conditions (existence of actuated artifacts, cases of permanent Korotkov's tone) results of oscillometric method with reference channel will be more accurate than .results of tacho-oscillographic method. Investigation with performing of BP measurement synchronically by different methods with existence of actuated artifacts and also comparison with measurement methods without cuff is promising in the future.